\documentclass{RE-modified-ws-mpla}

\newcommand{\be}{\begin{equation}}
\newcommand{\ee}{\end{equation}}
\newcommand{\ba}{\begin{eqnarray}}
\newcommand{\ea}{\end{eqnarray}}
\newcommand{\bit}{\begin{itemize}}
\newcommand{\eit}{\end{itemize}}
\newcommand{\eq}[1]{(\ref{#1})}

\renewcommand{\vec}[1]{\mbox{\boldmath $#1$}}
\newcommand{\evec}[1]{\mbox{\scriptsize\boldmath $#1$}}

\newcommand{\kvec}{\vec k}

\newcommand{\as}{\ensuremath{\alpha_s}}
\newcommand{\lqcd}{\Lambda_{\text{QCD}}}
\newcommand{\dsdt}{d\sigma/dt}

{\end{list}}

\usepackage{amsmath}

\begin{document}

\markboth{R.\ Enberg}
{Testing the Dynamics of High Energy Scattering}

%
\catchline{}{}{}{}{}
%

\title{TESTING THE DYNAMICS OF HIGH ENERGY SCATTERING USING VECTOR MESON PRODUCTION}

\author{\footnotesize RIKARD ENBERG\footnote{Present address:\ Service de Physique Th\'eorique,
CEA/Saclay, 91191 Gif-sur-Yvette Cedex, France (URA 2306 du CNRS)}
}

\address{Centre de Physique Th{\'e}orique, {\'E}cole Polytechnique, 91128 Palaiseau Cedex, France\footnote{Unit{\'e} mixte de recherche du CNRS (UMR 7644)}\\
enberg@cpht.polytechnique.fr}

\maketitle

\vspace{-42.8ex}
\hfill CPHT-RR-047.0804
\vspace{+41.8ex}


\begin{abstract}
I review work on diffractive vector meson production in photon--proton collisions at high energy and large momentum transfer, accompanied by proton dissociation and a large rapidity gap. This process provides a test of the high energy scattering dynamics, but is also sensitive to the details of the treatment of the vector meson vertex.

The emphasis is on the description of the process by a solution of the non-forward BFKL equation, i.e.\ the equation describing the evolution of scattering amplitudes in the high-energy limit of QCD. The formation of the vector meson and the non-perturbative modeling needed is also briefly discussed. 

\keywords{Vector meson production; diffraction; BFKL}
\end{abstract}

\ccode{PACS Nos.: 12.38.-t, 13.60.-r, 14.40.-n, 13.88.+e}

\section{Introduction}  

High energy reactions producing mesons, and in particular diffractive processes, are advantageous for studying various aspects of high energy scattering processes, partly because such processes are experimentally clean, and partly because many processes offer unique types of reactions. 

\begin{figure}[t]
\centerline{\epsfig{file=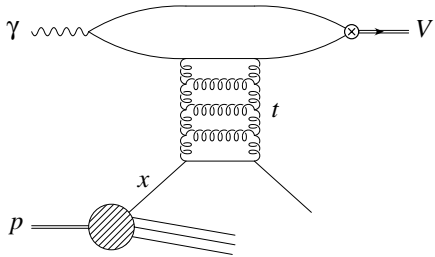,width=0.5\columnwidth}} %
\caption{Diffractive photoproduction $\gamma p \to V X$ of a vector meson at large momentum transfer.}\label{VMdiagram}
\end{figure}

This brief review is about a certain class of meson production, namely diffractive vector meson production in photon--proton collisions, where the color singlet $t$-channel exchange carries large momentum transfer, i.e., $\gamma p \to V X$ (see Fig.~\ref{VMdiagram}). In such processes there is a large rapidity gap centrally in the detector and the proton dissociates. The vector meson is on one side of the gap and on the other side of the gap there will be a jet, balancing the transverse momentum. This jet is not measured in experiments, but the cross section is obtained simply from the kinematics of the vector meson---a very nice experimental signal. The cross section is rather small, however, and the available HERA data\cite{ZEUSdata,H1data} is therefore very recent.

The mesons are identified experimentally by their decays into charged particles, e.g., $\rho \to \pi^+\pi^-$ or $J/\psi \to l^+l^-$, which provides a means of measuring the relative polarization by examining the angular distributions of the decay products.\footnote{The background to $\rho$ production is therefore the prompt production process $\gamma p \to \pi^+\pi^- X$, which could in principle be computed with the methods reported here by replacing the distribution amplitude of the $\rho$ meson by the generalized $\pi^+\pi^-$ distribution amplitude\protect\cite{Diehl}.}

This process is quite different from common diffractive processes where the proton survives intact and the rapidity gap is forward in rapidity. In general for photoproduction there can be two large momentum scales in the process---the squared momentum transfer $t$, and the vector meson mass $M_V$. In normal diffractive processes, $t=0$, and only large-mass meson production can be treated perturbatively. But if $-t\gg \lqcd^2$ perturbation theory can be used even for light mesons.

The $t$-channel exchange must be in a color singlet state and is 
often referred to as hard color singlet exchange, or hard pomeron exchange. 
The lowest order contribution is therefore the exchange of two gluons. 
At high center of mass energy $\sqrt s$, however, the amplitude is
dominated by terms proportional to $[\as \ln (-s/t)]^n$ and a fixed
order calculation is not adequate. Rather, one needs to resum
these logarithmic terms to all orders $n$. This is done by solving the 
Balitsky--Fadin--Kuraev--Lipatov (BFKL) equation\cite{BFKL}, which
resums the above logarithms in the leading logarithmic (LL)
approximation. 

The BFKL equation describes the evolution of QCD amplitudes in the high energy limit $s \gg -t$ and gives a new type of dynamics compared to fixed-order calculations. The resummation corresponds to considering diagrams with exchange of two gluons to all orders and keeping only those pieces containing a large logarithm. The main effects are that the two gluons are \emph{reggeized} due to virtual corrections, which means that the propagators acquire a Regge-like factor with a power of $-s/t$, and that any number of gluons is exchanged between the two gluons such that a \emph{gluon ladder} is formed.
This gives a cross section which goes as a power of the energy, like in Regge theory. See e.g.\ Refs.\ \refcite{Jeff-book,smallxcollaboration} for more on BFKL physics.

The status of the BFKL resummation framework is not yet clear because of uncertainties related to experimental limitations, small predicted effects at present accelerators,  NLL corrections, and so on. Many of the suggested signatures of BFKL resummation, such as the rise of $F_2$ at small $x$, can also be described by the traditional DGLAP equations. This makes it important to investigate possible signals for the behavior predicted by BFKL. In Ref.~\refcite{FR}, vector meson production at large $t$ was suggested to be one suitable process. Another large-$t$ process that has been advocated is large rapidity gaps between jets.\cite{CFL,EIM}

This review is organized around our calculations\cite{LVM1,LVM2} of vector meson production using BFKL resummation in the LL approximation.
We will make use of the non-forward version of the equation, which  has a complete analytical solution due to Lipatov.\cite{Lipatov} This solution, and our calculations, will be discussed in Section~\ref{BFKLsection}.

But there is of course also a lot of interest in the vector meson production process in itself. The experimental situation is not well understood from theory. The measured differential cross section $\dsdt$ for light meson production shows a $t$-dependence which goes roughly as $1/t^3$, and measurements of spin density matrix elements indicate that these mesons are predominantly transversely polarized; only 10--20\% of the cross section comes from longitudinally polarized mesons.
This is in contrast to the lowest order perturbative QCD calculations\cite{Ginzburg,Ivanov,GinzburgIvanov} that predict $\dsdt\sim 1/t^3$ for longitudinal mesons and $\dsdt\sim 1/t^4$ for transverse mesons---production of transversely polarized mesons is suppressed by an extra power of $t$! I will come back to these problems below and discuss various approaches to their solution. In particular I will show that our BFKL calculations give a natural explanation of these hierarchies.

\section{Review of the theory for vector meson production}

The process $\gamma p \to V X$ takes place by exchange of the hard pomeron as shown in Fig.~\ref{VMdiagram}. It has been shown\cite{pomtoquarks} that at large momentum transfer the pomeron couples predominantly to individual partons in the proton. One may therefore write the cross section as a product of the parton level cross section and the normal parton distribution functions of the proton,
\begin{align}
\frac{d\sigma (\gamma p \rightarrow VX)}{d t\, d x_j} \;=\;
\biggl(
\frac{81}{16}
G(x_j,t)+
\sum_{f}[q_{f}(x_j,t)+\bar{q}_{f}(x_j,t)]\biggr)\;
\frac{d\sigma (\gamma q \rightarrow V q)}{dt},
\label{dsdtgp}
\end{align}
where $G(x_j,t)$ and $q_{f}(x_j,t)$ are the gluon and quark
distribution functions respectively. The struck parton in the proton initiates a jet in the proton
hemisphere, and carries the fraction $x_j$ of the longitudinal momentum
of the incoming proton. We are working in the high energy regime, were the coupling of the pomeron to a quark is identical to the coupling to a gluon, apart from the appropriate color factor. This is the reason for the factor in brackets on the right hand side in Eq.\ \eq{dsdtgp}. So, in the following, all attention will be focused on the quasi-elastic subprocess $\gamma q \to V q$.

\subsection{Factorization}

The first computations\cite{Ginzburg} of vector meson photoproduction at arbitrary momentum transfer were done considering the Born graph for the pomeron, representing it by two gluons in a color singlet state. The results were given in the \emph{impact factor} representation of Cheng and Wu.\cite{ChengWu} This means that one writes the amplitude for the high energy process $A B\to C D$ on the form
\be
{\cal M}^{A B\to C D} (q) = 
\int {d^2 \vec k }
\frac{1}{\vec k ^2 (\vec q - \vec k)^2}
\varPhi^{A\to C}(\vec k, \vec q)\, \varPhi^{B\to D} (\vec k, \vec q),
\label{mpr}
\ee
where $\varPhi^{A\to C}$ and $\varPhi^{B\to D}$ are the impact factors for the upper and lower parts of the diagram, respectively.\footnote{For a derivation of this result and the expressions for the impact factors, see Ref.~\protect\refcite{Gavin}.} That is, they are the impact factors for the processes 
$A\to C$ and $B\to D$ with two gluons carrying transverse momenta $\vec k$ and $\vec q-\vec k$ attached, the gluons being in an overall color singlet state.

\begin{figure}[t]
\centerline{\epsfig{file=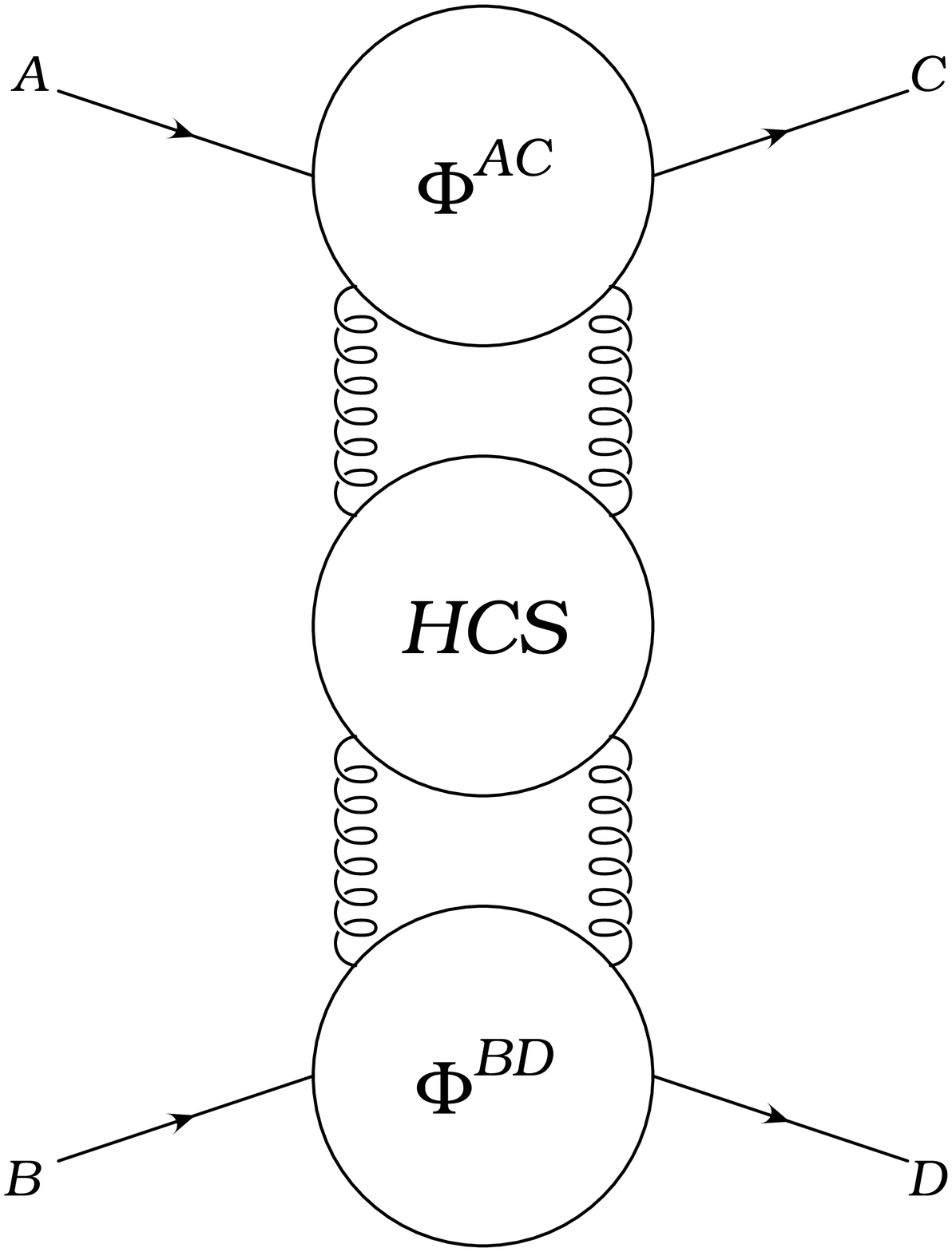,width=0.4\columnwidth}
\quad\raisebox{4cm}{\epsfig{file=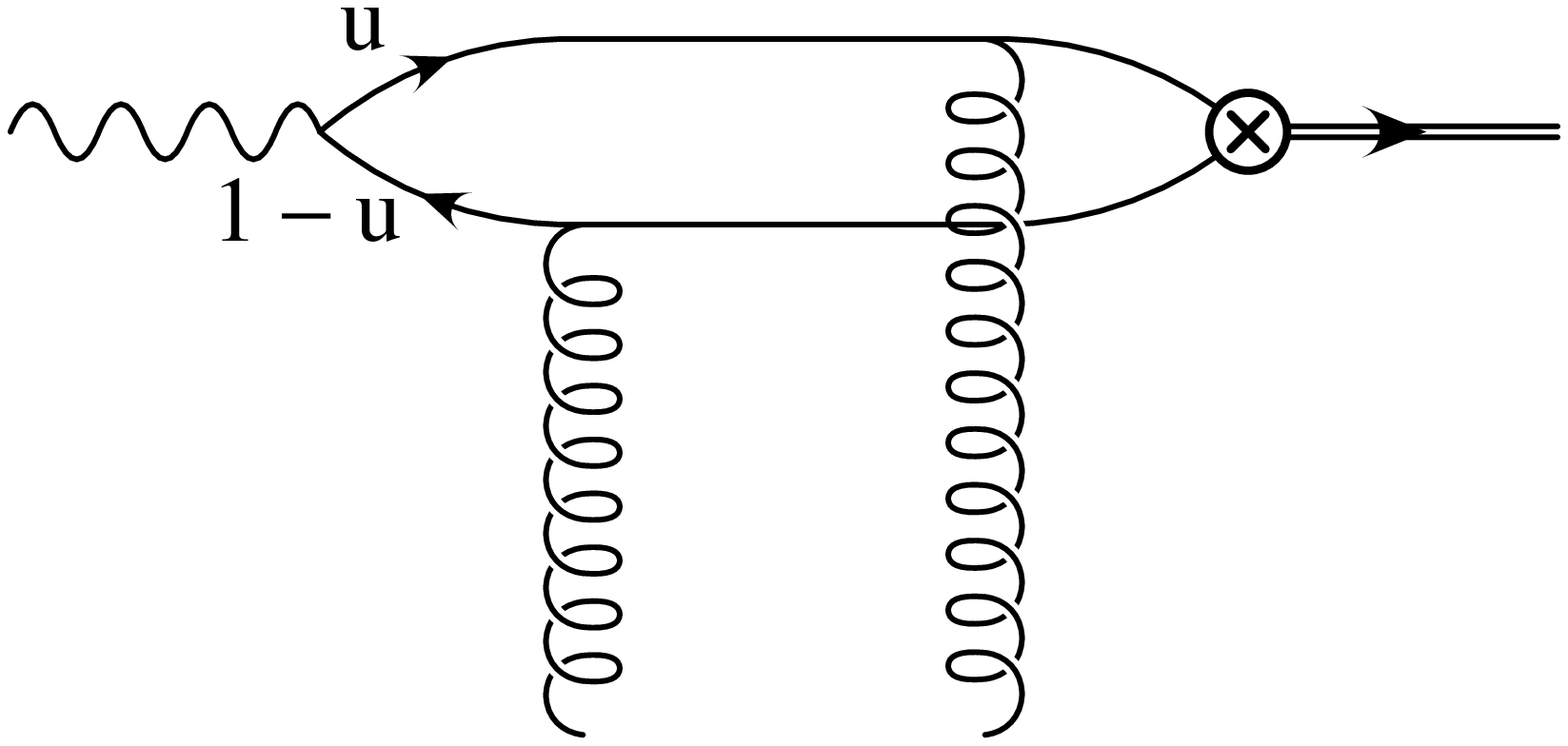,width=0.45\columnwidth}}}
\centerline{(a)\hspace{53mm}(b)\quad}
\caption{(a) $A B \to C D$ scattering process in impact factor representation. The upper and lower blocks represent the impact factors for the corresponding transitions with two gluons attached, and the middle block marked HCS (hard color singlet) represents the Green function for the two gluons. For pomeron exchange, this is the BFKL Green function. 
(b) The $\gamma\to V$ impact factor.}\label{impact}
\end{figure}
This factorization is schematically depicted in Fig.\  \ref{impact}a. It applies in the high-energy limit $s\gg -t$, where one can perform the integrals over the gluon plus and minus lightcone momenta, leaving only the integrals over the transverse components. The impact factorization simplifies calculations a great deal, because one can treat the process in three disjoint parts; the two impact factors up and down in the diagram, and the exchange mechanism in between. In the case of two-gluon exchange the middle block gives just delta functions connecting the gluon momenta and colors, and the denominator containing the propagators. In the BFKL case to be discussed later, the block contains the evolution of the two-gluon system into a gluon ladder, and will give a more complicated function.

In Fig.~\ref{impact}b, I show the lowest order diagram for the $\gamma\to V$ impact factor. The process is viewed in a frame where the photon has large lightcone momentum $q^+$ and fluctuates into a quark--antiquark pair before the hard scattering. This pair forms a color dipole which scatters by exchanging two gluons, and then forms a vector meson, which is possible because for pomeron exchange the vector meson has the same quantum numbers as the photon. The fluctuation of the photon into the dipole is long-lived, and the transverse size of the pair is frozen during the scattering.

Also here there is a factorization formula which applies in the diffractive limit of negligible quark lightcone minus-momentum. The impact factor is given by
\be
\varPhi_{\gamma(\lambda) V(\lambda')} (\vec k,\vec q) \sim
\int d u \int d^2 \vec r \;
\varPsi^* _{V(\lambda')} (u,\vec r) \,
T (\vec k, \vec q; \vec r,u)\,
\varPsi_{\gamma(\lambda)}(u,\vec r),
\label{fact01}
\ee
where $\varPsi_{V}$ ($\varPsi_\gamma$) denotes the meson (photon) wave function (which will be discussed below), $\lambda$ ($\lambda'$) denotes the polarization of the photon (meson),
$u$ is the quark
lightcone plus-momentum fraction of the meson momentum,
$\vec r$ is the transverse quark--antiquark separation, 
and 
$T(\vec k, \vec q; \vec r,u)$ is the hard amplitude of the QCD dipole scattering, which, 
up to a constant factor, equals
$f^{\text{dipole}} =
e^{i{\evec q}{\evec r}u}\left( 1 - e^{-i{\evec k}{\evec r}} \right)
\left(1 -  e^{-i({\evec q-\evec k}){\evec r}}   \right).$
The impact factor for the $q\to q$ transition is, in the high energy limit (where one uses the eikonal approximation), just a constant.\footnote{There are some subtleties regarding the pomeron coupling to colorful particles. This was addressed by Mueller and Tang\protect\cite{MT}. See also the discussion in Refs.\ \protect\refcite{pomtoquarks} and \protect\refcite{MMR}.}

The above results are given in a mixed representation in transverse momenta for the gluons and transverse vectors in coordinate space for the quarks. They can of course be written in the more familiar momentum space by a suitable Fourier transform, but the present form turns out to be quite useful for BFKL calculations.

\subsection{The photon and meson wave functions}

The photon and meson wave functions give the probability amplitudes for the splittings $\gamma\to q\bar q$ and $V\to q\bar q$. The photon wave function can be straightforwardly calculated in QED (see e.g.\ Refs.~\refcite{Jeff-book,BrodskyVM}) and I do not reproduce the result here.\footnote{Though, when calculating the impact factor one might instead calculate the hard scattering amplitude for the whole $\gamma g g \to q\bar q$ process directly and convolute it with the meson wave function. This is actually what is done in most of the literature cited here.}

The meson wave function is more complicated. It contains non-perturbative information on the transition of the quark--antiquark pair into a vector meson;  thus a model is needed. This model will usually involve the \emph{distribution amplitudes} (DAs) of the meson. Roughly speaking, the DA $\phi(u)$ describes the distribution of longitudinal momentum fraction $u$ of the quark in the meson. 
In the early papers on light meson production\cite{BrodskyLepage,BrodskyVM} their form was not known, but one used either the asymptotic form $\phi(u)=6u(1-u)$, or a form suggested by Chernyak and Zhitnitsky\cite{CZ} based on QCD sum rules. These calculations have invariably set the quark mass to zero.
For heavy mesons, on the other hand, a non-relativistic quarkonium picture of the meson has been used, where the quark and the antiquark have the same longitudinal momentum, $\phi(u)=\delta(u-1/2)$.
In the BFKL calculations\cite{LVM1,LVM2} to be described below, we unify light and heavy mesons by keeping the quark masses throughout. We are therefore able to describe all mesons within the same framework.

From quite general considerations 
one finds that the meson wave function $\varPsi_{V}$ above is related to the meson-to-vacuum transition matrix elements\footnote{See Refs.~\refcite{IKSS} and \refcite{Gavin} for more details.}
$\langle V(p,\lambda) | \bar\psi(-r/2) \varGamma^t \psi(r/2) | 0 \rangle$ where $\varGamma^t$ are Dirac-matrix structures from the Fierz decomposition of the amplitude. These matrix elements are precisely those that appear in the lightcone definitions of the DAs and have been calculated from QCD sum rules with the inclusion of higher twists,\cite{BBKT,BB} extending the results of Ref.~\refcite{CZ}. I will not list the expressions here, suffice it to say that the distribution amplitudes of the vector mesons appear, and that they allow the calculation of the production amplitudes for all meson polarizations and quark helicities.

\subsection{Light meson calculations}\label{lightmesons}

The hard scattering amplitude and impact factors were calculated quite some time ago in lowest order QCD by Ginzburg, Panfil, and Serbo,\cite{Ginzburg} who obtained the amplitudes for production of \emph{longitudinally} polarized light mesons in photon--photon and photon--quark collisions. Ginzburg and Ivanov\cite{GinzburgIvanov} added transversely polarized mesons to the calculation, and Ivanov\cite{Ivanov} extended it to include BFKL effects.

These results, however, have the features mentioned in the Introduction:\ the longitudinal mesons dominate the cross section for light mesons---because of the chiral even nature of the quark--photon coupling for massless quarks, the production of transversely polarized mesons vanishes as the quark mass tends to zero.

A solution to this problem was suggested by Ivanov, Kirschner, Sch\"afer, and Szymanowski,\cite{IKSS} who argued that the photon splitting may also have a non-perturbative chiral odd component\footnote{The chirality of the $q\bar q$ state refers to the spinorial structure of the wave function---upon Fierz decomposing the amplitude, one gets terms with definite chiral parity. The chiral even terms are proportional to $\gamma^\mu$ and $\gamma^\mu\gamma_5$ and are non-zero in the massless quark limit; the chiral odd terms are proportional to $\sigma^{\mu\nu}$ and vanish in the massless limit. These cases correspond to the quark and antiquark having antiparallel (chiral even) or parallel (chiral odd) helicities.} which would enhance the production of transverse mesons.
This chiral odd photon wave function was 
modeled on a meson-like wave function, with a coupling constant arising from the quark condensate $\langle \bar q q \rangle$. The coupling constant is proportional to $\langle \bar q q \rangle$ and the magnetic susceptibility of the vacuum, which were taken from QCD sum rules estimates. The chiral symmetry breaking therefore occurs also when neglecting quark masses. 

This approach enhances the production of transverse mesons, such that it numerically dominates in the region of moderate momentum transfer that is experimentally observed, but longitudinal meson production still dominates for asymptotically large momentum transfers. However, the transverse meson cross section still scales as $1/t^4$ and the results do not agree with experimental data\footnote{See Ref.\ \refcite{IK} for a comparison of the results of Ref.\ \refcite{IKSS} to the ZEUS data\cite{ZEUSdata}.}. Furthermore, this calculation used two-gluon exchange and will give a flat energy behavior.

Chiral odd components were also included by Enberg, Forshaw, Motyka, and Poludniowski together with complete BFKL evolution in Refs.~\refcite{LVM1} and \refcite{LVM2}, but with a different non-perturbative treatment of the chiral odd amplitudes. As we will see this gives a $t$-dependence of the differential cross section which agrees very well with data. Moreover, transversely polarized mesons dominate for moderate momentum transfers. (See Section~\ref{BFKLsection} for more details).

In yet another approach, Ivanov and Kirschner\cite{IK} calculated the process using a meson wave function with explicit $\kvec$-dependence obtained from the photon wave function through the use of QCD sum rules. This meson wave function reduces to the distribution amplitude when integrated over transverse momentum, and allows studying the effects of large quark--antiquark dipoles, since these dipoles have been assumed small in other analyses.
The calculation used Born level pomeron exchange, but additionally included double logarithmic corrections as an approximation of the BFKL evolution. Chiral odd contributions were not included.

\subsection{Heavy meson calculations}

Heavy meson production ($J/\psi$ and $\varUpsilon$) is usually considered in the approximation of very heavy quarks, in which the quarks have zero transverse momentum and share the longitudinal momentum of the meson equally. The first BFKL calculation of this process was done by Forshaw and Ryskin,\cite{FR} who solved the BFKL equation approximately. Bartels, Forshaw, Lotter and W\"usthoff\cite{Bartels} then computed the scattering amplitude analytically for asymptotically large rapidities using Lipatov's BFKL solution. This result was compared to HERA data by Forshaw and Poludniowski\cite{FP}, who found good agreement between theory and data for heavy mesons. But these authors also considered data on $\rho$ and $\phi$ production. Surprisingly, the predicted differential cross sections agreed quite well with light meson data, although the heavy quark approximation should not be applicable for these mesons. This agreement is understood when considering the full BFKL amplitudes,\cite{LVM1,LVM2} as I will explain below.

\begin{figure}[t]
\centerline{\epsfig{file=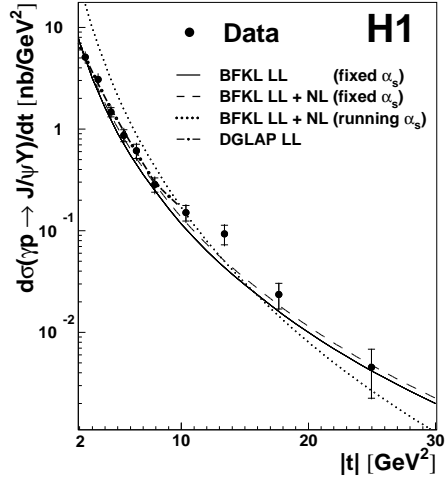,width=0.5\columnwidth}} %
\caption{BFKL computation\protect\cite{HVM} compared to H1 data\protect\cite{H1data} on the differential cross section. The solid line shows the LL BFKL result mentioned in the text. This plot is taken from Ref.~\protect\refcite{H1data}.}\label{Jpsifig}
\end{figure}

The above calculations were all done in the approximation of asymptotically large energies (i.e.\ rapidities). Technically this amounts to retaining only the leading \emph{conformal spin} in the expansion of the full BFKL amplitude. The conformal spin is an integer $n$, which labels the eigenfunctions
and the full amplitude is a sum over all $n$. The only term in the sum that does not vanish in the limit of rapidity $y\to\infty$ is $n=0$, so to simplify the calculations the other terms are often dropped. However, it turned out\cite{EIM,MMR} that to describe the cross section for gaps between jets using BFKL, it was necessary to keep the subleading terms in the sum. As a consequence, Enberg, Motyka, and Poludniowski\cite{HVM} computed the analytical expression for the heavy meson production amplitude for all conformal spins. It was found that for $J/\psi$ production the non-zero conformal spins contribute roughly 10\% to the cross section at the HERA experiments. This calculation has been compared to recent data with good agreement, see Fig.\ \ref{Jpsifig}.

\section{General BFKL result for vector meson production}\label{BFKLsection}

This section contains a review of the work reported in Refs.\ \refcite{LVM1,LVM2} mentioned above. The purpose of these papers was to compute the complete set of amplitudes for the vector meson production process $\gamma q\to V q$ by BFKL pomeron exchange, for any momentum transfer $t$, any meson polarization $\lambda$, and any vector meson $V = \rho, \phi, \omega, J/\psi, \varUpsilon$, unifying and generalizing previous work. 

I have already stressed above that due to large logarithms, BFKL evolution of the pomeron exchange is needed in order to obtain the correct Regge-type energy dependence of the amplitudes\footnote{See e.g.\ the comparison with data in Ref.\ \protect\refcite{H1data}.}. We will also see that the BFKL resummation has the effect that amplitudes are always finite, even for zero quark masses---in contrast to the Born level two-gluon exchange. Furthermore, the resummed amplitudes are dominated by configurations where the pomeron momentum is shared between the two gluons, instead of very asymmetrically distributed as for two-gluon exchange.

In summary, the calculated amplitudes have the features that 
(i)
the pomeron exchange has LL BFKL evolution, for all conformal spins; 
%
(ii)
quark masses are included for all vector mesons;
%
(iii)
chiral odd photon splittings are considered by using the constituent quark mass in the perturbative QED wave function; 
%
(iv)
full higher twist distribution amplitudes are taken into account, and both the asymptotic and non-asymptotic forms have been tried; 
and
%
(v)
all amplitudes are finite at the endpoints ($u\to 0,1$).
Thus, the results in this section are very general and contain all the earlier calculations\cite{Ginzburg,IKSS,FR,Bartels,FP,HVM} in various appropriate limits.

\subsection{Impact factors}

We want to compute the helicity amplitudes for the vector meson process, ${\cal M}_{\lambda \lambda'}$, where $\lambda$ and $\lambda'$ refer to the polarization of the incoming photon and of the outgoing vector meson. Since the photon is real, we need only consider transversely polarized photons, which we can take to be $+$ polarized. The meson has $\lambda'=+,-$ or $0$, which means that we need to compute the helicity amplitudes ${\cal M}_{++}$, ${\cal M}_{+0}$  and ${\cal M}_{+-}$.

Because the amplitude is factorized, we may start with the amplitude for two-gluon exchange according to Eq.\ \eq{mpr}, and then replace the middle block in Fig.\ \ref{impact} by the BFKL pomeron instead of just the two gluons. According to Eq.\ \eq{mpr}, we then need the impact factors
$\varPhi^{\gamma(\lambda)\to V(\lambda')}$ and $\varPhi^{q\to q}$ for the upper and lower blocks in Fig.\ \ref{impact}. The quark impact factor is a constant, and we choose the normalization $\varPhi^{q\to q}=1$, putting all non-trivial factors into the photon--meson impact factor. 

We include both the chiral even and chiral odd photon splittings, writing
$\varPhi = 
\varPhi_{\text{even}} + 
\varPhi_{\text{odd}}$ for all impact factors. The chiral even impact factors were computed in Refs.\ \refcite{Ginzburg,IKSS} for massless quarks. Instead of the non-perturbative photon wave function of Ref.\ \refcite{IKSS} we account for the non-perturbative coupling and chiral symmetry breaking by calculating the chiral odd impact factors with a perturbative QED photon wave function, replacing the current quark mass with its constituent mass. Note that all the chiral odd amplitudes vanish for vanishing quark mass.

Full expressions for the impact factors for all helicities and chiralities are given in Ref.\ \refcite{LVM1}. Here, as an example, I will only display one of them,
\ba
\varPhi^{\gamma(+) \to V(+)} _{\text{even}} = &&
i \alpha_s ^2 \frac{N_c^2 -1}{N_c^2} e Q_V
\int \frac{d^2{\vec r}\;d u}{4\pi}\;
m |\vec r|\, K_1 (m |\vec r|)
\,(\vec \epsilon^{(+)}\cdot\vec \epsilon^{(+)*})
\nonumber \\
&&
\times 
f^{\text{dipole}}
f_V M_V
\frac{u{(1-u)}}{2}\left(\int\limits_0^u \,
\frac{d v}{{(1-v)}}\,\phi_{||}(v) +
\int\limits_u^1 \,\frac{d v}{v}\,
\phi_{||}(v)   \right),
\label{if++}
\ea
where $m$ is the quark mass, and $M_V$, $f_V$ and $Q_V$ are the mass, decay constant and effective quark charge of the vector meson, respectively. $\phi_{||}(u)$ is the twist-2 distribution amplitude for the vector meson.

\subsection{Solution of the BFKL equation}

The next task is to replace the two-gluon exchange mechanism with the full BFKL pomeron. In order to do this, we use the beautiful solution of the BFKL equation by Lipatov,\cite{Lipatov} which was obtained drawing on the conformal symmetry of the BFKL kernel when expressed in two-dimensional complex gluon coordinate space. The amplitude of a scattering process takes the form of an expansion in the basis of conformal eigenfunctions $E_{n,\nu}$ of the BFKL kernel,
\ba
{\cal M} \;=\;
\sum_{n}
\int
d\nu\; \frac{(\nu^{2}+n^{2}/4) \; \exp [(3\as/\pi) \chi_{n}(\nu) y]}{[\nu^{2}+(n-1)^{2}/4][\nu^{2}+(n+1)^{2}/4]}
\; \left( E_{n,\nu} |\varPhi_{1} \right)    
\left( \varPhi_{2} | E_{n,\nu} \right),    
\label{bfklampl}
\ea
where $n$ is the conformal spin, $y=\log(\hat s/\Lambda)$ is the rapidity, and $\chi_{n}(\nu)$ is the characteristic BFKL function given by
$\chi_{n}(\nu)= 2 \text{Re} \left(\psi(1)-\psi(1/2+|n|/2+i\nu)\right)$. The factor $\left( \varPhi_{1} | E_{n,\nu} \right)$ is a schematic way of representing the projection of the impact factor on the eigenfunction, which is  given by
\begin{equation}
\left( \varPhi_{1} | E_{n,\nu} \right)\; = \; \frac{1}{2\pi} 
\int \frac{d^{2}k}{(2\pi)^{2}}
\varPhi_{1}(k,q)\int
d^{2}\rho_{1}\,d^{2}\rho_{2}\; E_{n,\nu}(\rho_{1},\rho_{2})
\exp (i k\cdot \rho_{1} +i(q-k) \cdot \rho_{2}),
\label{impf}
\end{equation}
where the eigenfunctions are 
\begin{equation}
E_{n,\nu}(\rho_{1},\rho_{2})\;=\;
\biggl(\frac{\rho_{1}-\rho_{2}}{\rho_{1}\rho_{2}}\biggr)^{-\widetilde\mu+1/2}
\biggl(\biggl(\frac{\rho_{1}-\rho_{2}}{\rho_{1}\rho_{2}}\biggr)^{*}
\biggr)^{-\mu+1/2},
\label{E}
\end{equation}
with $\mu=n/2-i\nu$ and $\tilde{\mu}=-n/2-i\nu$.
Here $k$ and $q$ are transverse two dimensional momentum vectors, and
$\rho_{1}$ and $\rho_{2}$ are  space vectors in the complex representation.

Computing all integrals that can be done analytically, one finds that all six helicity amplitudes take the form
\begin{eqnarray}
&& {\cal M}_{\lambda\lambda'}^{\text{even/odd}} = \frac{1}{|q|}\,
\sum_{n=-\infty}^{n=+\infty}
\int_{-\infty}^{\infty}d\nu
\int_0 ^1 d u\; \varPhi_{\lambda\lambda'}^{\text{even/odd}}(u)\,\label{bfklamplitude} 
\\
&& \times 
\frac{\nu^{2}+n^{2}}{[\nu^{2}+(n-1/2)^{2}][\nu^{2}+(n+1/2)^{2}]}
\frac{\exp [(3\as/\pi) \chi_{2n}(\nu) y]}{\sin (i\pi\nu)} \,
I_{\alpha \beta}(\nu,2n,q, u;a).
\nonumber
\end{eqnarray}
Here, $\varPhi_{\lambda\lambda'}^{\text{even/odd}}(u)$ are functions (in which all constants have been absorbed) which are combinations of the distribution amplitudes, and which specify the $u$-dependence for different helicity combinations.  $I_{\alpha \beta}(\nu,2n,q, u;a)$  is given by a line integral,
\ba
I_{\alpha\beta}(\nu,n,q,u;a) &=&
\frac{m}{2}\int^{C^{\prime}+i\infty}_{C^{\prime}-i\infty}
\frac{d\zeta}{2\pi i}\varGamma(a/2-\zeta)\varGamma(-a/2-\zeta)\,
\tau^{\zeta} \; (i\, \text{sign}\,(1-2u))^{\alpha-\beta+n}  \nonumber \\
&\times&  \left(\frac{4}{|q|}\right)^{4}
\left[\sin\pi(\alpha + \mu + \zeta)\; \right.
B(\alpha,\mu, q^* ,u,\zeta)\,
B(\beta,\widetilde\mu,q,u^* ,\zeta) \nonumber \\
&-& (-1)^n
\sin\pi(\alpha - \mu + \zeta)\;
B(\alpha,-\mu, q^* ,u,\zeta)\,
B(\beta,-\widetilde\mu,q,u^* ,\zeta)
\left. \right],
\label{finalintegral}
\ea
where we have introduced the dimensionless parameter $\tau = 4 m^2/|q|^2$
and the conformal blocks
\ba
B(\alpha,\mu, q^* ,u,\zeta) =  
(-4u (1-u))^{-(\mu+2+\alpha+\zeta)/2}
\left(\frac{4}{ q^* }\right)^\alpha
2^{-\mu}\,
\frac{\varGamma(\mu+2+\alpha+\zeta)}{\varGamma(\mu+1)} \nonumber \\
_2F_1\left(\frac{\mu+2+\alpha+\zeta}{2} \, , \,
\frac{\mu-1-\alpha-\zeta}{2}\, ; \,
\mu+1\, ; \,\frac{1}{4u (1-u)}\right).
\label{blocks2}
\ea
The parameters $\alpha$, $\beta$, and $a$ above take on different constant values for different amplitudes 
(see Ref.\ \refcite{LVM1} for a detailed derivation of formulas (\ref{finalintegral}, \ref{blocks2})).

The nice symmetric structure of these expressions is a consequence of the original conformal symmetry of the BFKL kernel. Note that the functions $B$ are not single-valued as they stand, but that the complete integrand in Eq.\ \eq{finalintegral} is. The sum over conformal spin includes all terms with even $n$, although they are rapidly becoming very small and only about a dozen are needed for phenomenological applications. However, only the terms with odd $n$ vanish identically, in contrast to the similar case when the vector meson is replaced by a photon.\cite{Evanson,Munier} In this case \emph{only} the terms with $n=0,\pm 2$ contribute and the others vanish. The reason for this is not known.

\subsection{Properties of the solutions}

The BFKL resummation, in contrast to the two-gluon approximation, makes the amplitudes infrared finite in the endpoint region. In Ref.\ \refcite{IKSS} the amplitudes ${\cal M}_{++}^{\text{even}}$, ${\cal M}_{+0}^{\text{odd}}$, and ${\cal M}_{+-}^{\text{odd}}$ were found to have divergent integrals over $u$, while in the BFKL case these integrals converge. One may think that this regulation of the divergence is due to the inclusion of quark masses, but in fact the amplitudes of Ref.\ \refcite{LVM1} remain convergent in the limit $m\to 0$; in Appendix B of Ref.\ \refcite{LVM2} it is shown that the amplitudes converge unless one takes the simultaneous limit $m\to 0$ and $y\to 0$. 

In Ref.\ \refcite{IKSS} it is suggested that the endpoint divergences could be suppressed by a Sudakov factor limiting radiation off the soft dipole configurations. In Ref.\ \refcite{Hoyer} on the other hand, it is argued that it is precisely these endpoint divergences that provide the solution to the problem that the amplitudes do not scale with $t$ as expected. The divergence is taken as a sign that factorization breaks down, and treating this correctly provides an extra factor of $t$. For more details, I refer to this paper. 

\begin{figure}[t]
{\epsfig{file=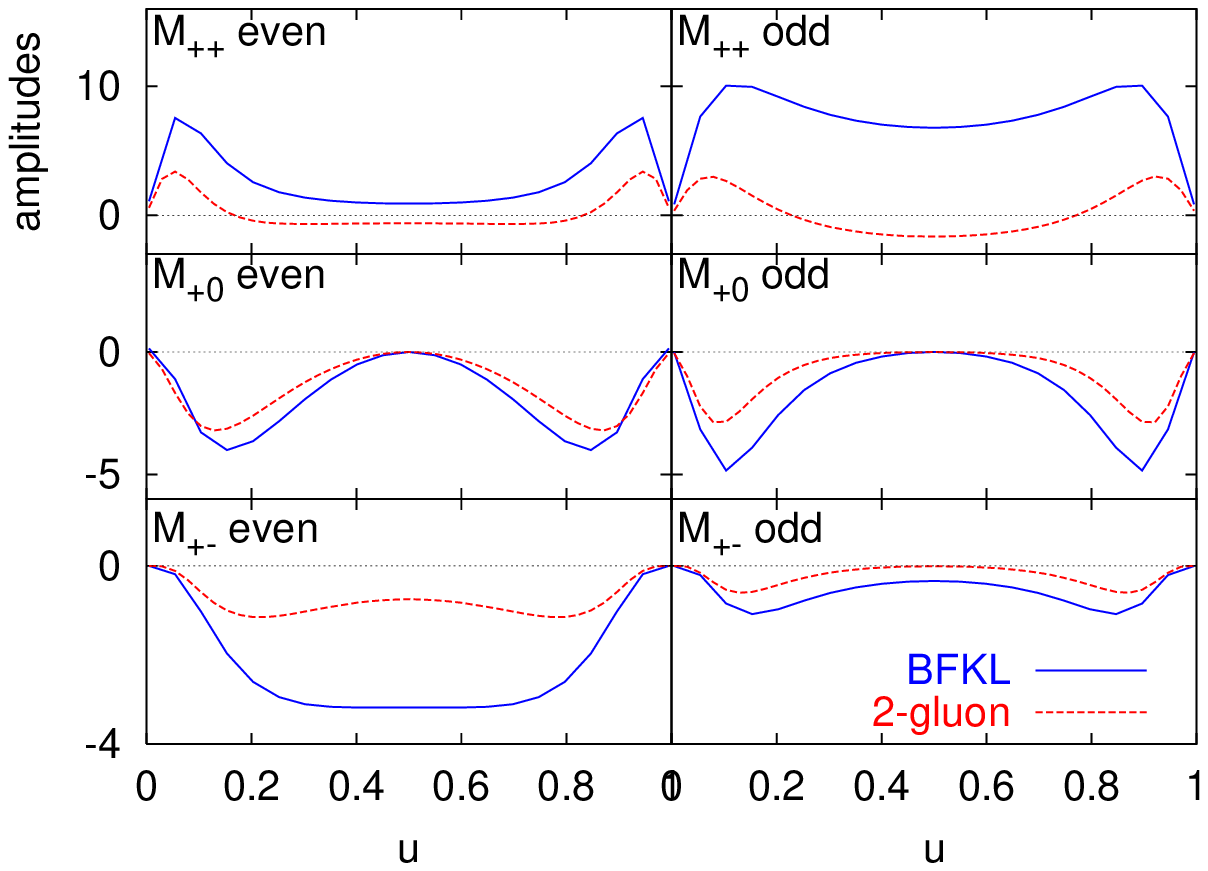,width=0.48\columnwidth,clip=}}
{\raisebox{5cm}{\epsfig{file=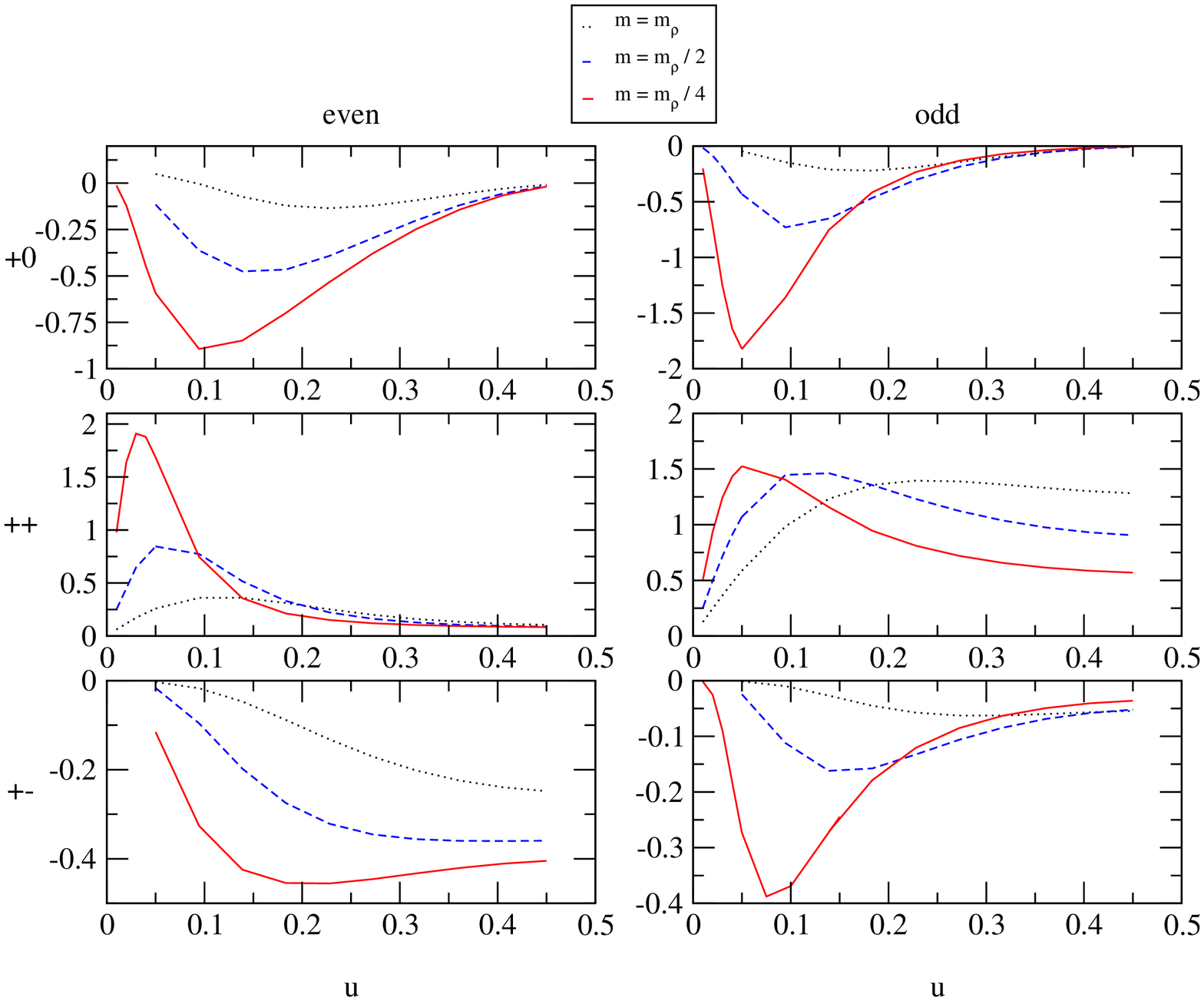,angle=-90,width=0.48\columnwidth,clip=}}\\
\centerline{(a)\hspace{0.45\columnwidth}(b)}} 
\caption{Helicity amplitudes differential in $u$, at fixed $y$ and $t$. (a) BFKL results compared to the massive quark two-gluon result. The two-gluon amplitudes have been multiplied by a factor three. (b) BFKL results for three different constituent quark masses, $m=M_V/2$ (blue, dashed curves in the middle), and $m=M_V/4$ and $m=M_V$. From Ref.\ \protect\refcite{LVM2}.}\label{udepend}
\end{figure}

The regulation of the endpoint divergences begs the question of how the
$u$-dependence of the amplitudes is affected by BFKL effects. 
Fig.\ \ref{udepend}a shows the six helicity amplitudes differential in $u$ calculated from BFKL and from two-gluon exchange including quark masses, after evaluating the remaining integrals numerically. As discussed above the quark masses regulate the endpoint divergence of the two-gluon amplitudes, so the divergences are not seen in this plot, but it shows that BFKL somewhat suppresses the contributions of asymmetrical dipoles relative to the two-gluon curves. 

But how is this related to the choice of constituent quark mass? The physically motivated choice for the quark mass is $m=M_V/2$, but in our calculation this is a non-perturbative parameter which may be adjusted. The answer is that when the constituent quark mass is increased the curves are more concentrated towards $u=1/2$, see Fig.\ \ref{udepend}b. This is especially true for the amplitudes that were divergent for the massless two-gluon case, ${\cal M}_{++}^{\text{even}}$, ${\cal M}_{+0}^{\text{odd}}$, and ${\cal M}_{+-}^{\text{odd}}$, which seem to have some memory of said divergences. 

Finally, with the complete BFKL amplitudes we can investigate how the heavy quark approximation used earlier\cite{FR,Bartels,FP,HVM} arises from the more general amplitudes. This is studied in some detail in Ref.\ \refcite{LVM2}. As was discussed above, the approximation consists in taking the quarks to be very heavy and to move together without relative motion. This means that their longitudinal momenta are equal, leading to the replacement $\phi(u) \to \delta(u-1/2)$. Moreover, the transverse momenta of the quark and antiquark are zero, which in coordinate space means that the transverse size of the pair is zero. These two replacements leave only the ${\cal M}_{++}^{\text{odd}}$ amplitude non-zero. But in fact, it is precisely the ${\cal M}_{++}^{\text{odd}}$ amplitude that dominates the cross section also for the full BFKL result. The obvious conclusion is that the reason for the phenomenological success\cite{FP} of the heavy quark approximation (even for light mesons where there is no reason to expect it to work) is due to the selection of the dominant amplitude in that approximation. This success, however, is limited, because there is no helicity flip of the quarks, and the mesons produced are all transversely polarized, at variance with experimental data.\cite{ZEUSdata,H1data}

\subsection{Comparison with data}

Now has come the time to consider how the theoretical amplitudes compare to the experimental data. In Ref.\ \refcite{LVM2} we performed a thorough comparison of both the BFKL and two-gluon results calculated in Ref.\ \refcite{LVM1} with ZEUS data\cite{ZEUSdata} on $\rho$ and $\phi$ production and H1 data\cite{H1data} on $J/\psi$ production. In this section I will therefore only highlight a few of the results and conclusions.

In the leading logarithmic approximation it is not prescribed how to treat the running coupling $\as$, and the normalization of the cross sections is therefore not completely fixed. The running coupling actually appears in two distinct places, in the coupling of the two gluons to the impact factors and in the coupling of gluons inside the BFKL ladder. We use a fixed value for the latter,\footnote{This is well-motivated and common practice in LL phenomenology. Furthermore, it has been argued\protect\cite{Brodsky} that the effect of higher order logarithms is to effectively freeze the coupling.} but we use both fixed and running $\as$ in the prefactor. The definition of the energy scale in the rapidity, $y=\ln (s/\Lambda^2)$, is not prescribed at our level of accuracy either, and we use $\Lambda^2=M_V^2-t$. Thus, in the end, we have three parameters in our fits, the two $\as$ and the quark mass $m$, chosen as $m=M_V/2$ in comparing to the data.

\begin{figure}[t]
\centerline{\epsfig{file=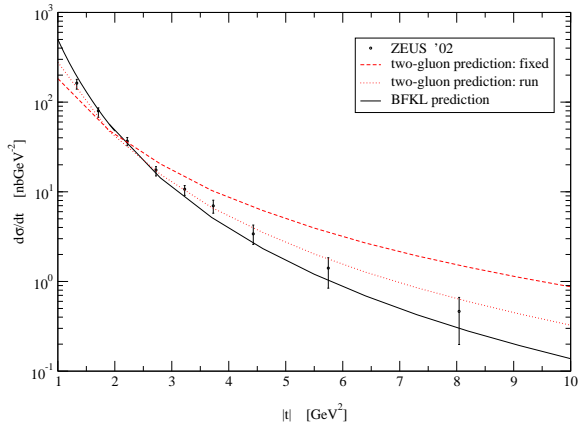,angle=-90,width=0.6\columnwidth,clip=}} %
\caption{Differential cross section for $\rho$ photoproduction:\ theory compared to ZEUS data\protect\cite{ZEUSdata}. The solid line shows the BFKL result with fixed coupling $\as^{IF}=0.17$ and $\as^{BFKL}=0.25$. The dashed and dotted lines show the two-gluon result with fixed and running coupling, respectively.}\label{dsdt-rho}
\end{figure}

\begin{figure}[t]
\centerline{\epsfig{file=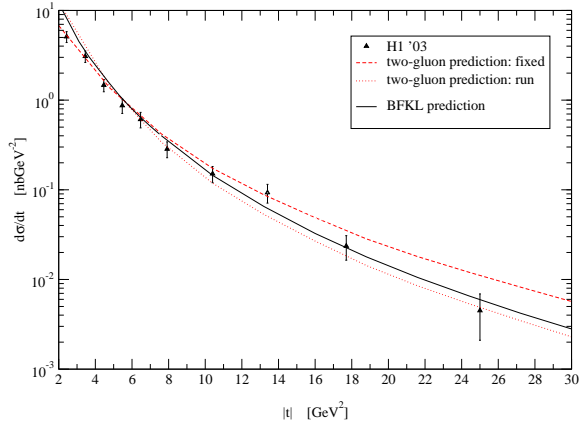,angle=-90,width=0.6\columnwidth,clip=}} 
\caption{Same as Fig.\ \protect\ref{dsdt-rho} but for $J/\psi$ production and H1 data\protect\cite{H1data}.}\label{dsdt-jpsi}
\end{figure}

The BFKL differential cross sections $d\sigma/dt$ are found to be in very good agreement with the data, see Fig.\ \ref{dsdt-rho} for $\rho$ and \ref{dsdt-jpsi} for $J/\psi$ production. The same parameters have been used for all mesons (including $\phi$, which is not shown here). We see that $J/\psi$ can be reproduced with the full BFKL calculation using the DAs, in addition to the earlier calculations in the heavy quark approximation shown in Fig.\ \ref{Jpsifig}. The two-gluon curve with running coupling is in as good agreement as the BFKL curve, but with fixed coupling it is clearly too flat. The reason is that the flatter ${\cal M}_{+0}$ amplitude is much larger in the two-gluon approximation than in the BFKL case.

\begin{figure}[t]
\centerline{\epsfig{file=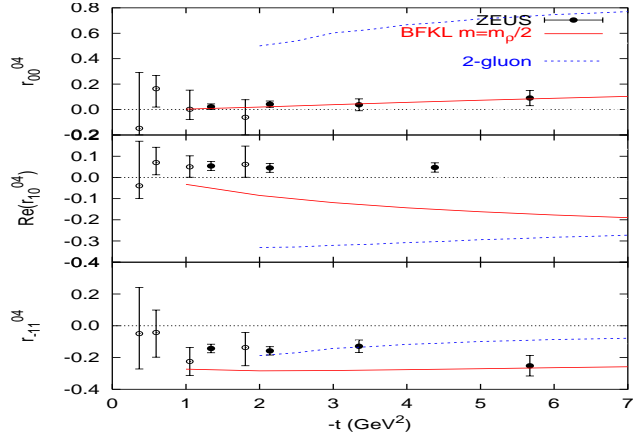,width=0.65\columnwidth,height=0.45\columnwidth,clip=}} %
\caption{Spin density matrix for $\rho$ photoproduction:\ theory compared to ZEUS data\protect\cite{ZEUSdata}. The solid line shows the BFKL result and the dashed line shows the two-gluon result.}\label{rij-rho}
\end{figure}

This is evident when looking at the measured spin density matrix elements $r^{04}_{ij}$, which are dimensionless combinations of the different ${\cal M}_{\lambda\lambda'}$ amplitudes. They can be measured experimentally through  angular distributions of the meson decay products. This provides the only way of studying the helicity structure of the process; to measure individual helicity cross sections one would need polarized beams.

There are three relevant spin density matrix elements in this process:\
$r^{04}_{00} \propto |{\cal M}_{+0}|^2$, 
$r^{04}_{10} \propto \text{Re}[{\cal M}^*_{+0}({\cal M}_{++}-{\cal M}_{+-})]$, and $r^{04}_{-11}\propto \text{Re}[{\cal M}_{++}{\cal M}_{+-}^*]$. All are normalized to the sum of all amplitudes. $r^{04}_{00}$ directly gives the ratio of the ${\cal M}_{+0}$ to the full amplitude and as seen in Fig.\ \ref{rij-rho} for $\rho$ production, this is well reproduced by BFKL but completely wrong for the two-gluon curve. Thus, even though the cross section was described by two-gluon exchange, the dynamics is not correct and the dominance of transverse mesons seen in data is not reproduced.  

However, we also find that $r^{04}_{-11}$ is somewhat too negative for the BFKL curve, while the real failure shows up for $r^{04}_{10}$: both the BFKL and two-gluon curves have the wrong sign! This shows that the ${\cal M}_{+0}$ amplitude has the wrong sign in both calculations, and consequently that it is not under control. There are large end point contributions in the calculation, which may, as in Ref.\ \refcite{Hoyer}, signal that the used factorization is breaking down.

\begin{figure}[t]
\centerline{\epsfig{file=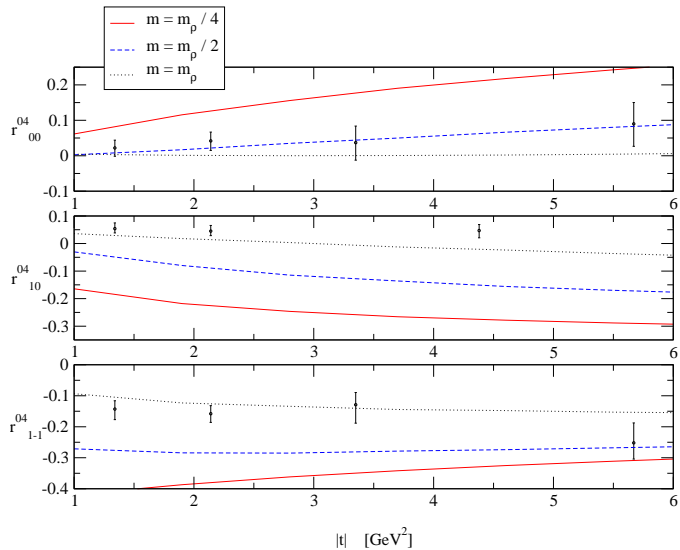,angle=-90,width=0.7\columnwidth,clip=}} 
\caption{Helicity matrix elements from BFKL with three different choices of constituent quark mass, $m=M_V/4$ (solid line), $m=M_V/2$ (dashed line), and  $m=M_V$ (dotted line).}\label{rij-mass}
\end{figure}

An interesting observation, however, is that if the quark mass is increased to double its physically motivated value, then the sign of $r^{04}_{10}$ is switched to the correct one (the cross section is much less sensitive). This is illustrated in Fig.\ \ref{rij-mass}. This increase of the mass may be somewhat artificial, but it serves to show an interesting point:\ Increasing the quark mass effectively leads to a stronger suppression of large dipoles (through the Bessel functions in the impact factors). There are several possible sources of such a suppression:\ saturation effects, Sudakov suppression of radiation off the quark and antiquark, or inclusion of the size of the vector meson in the wave function such as in Ref.\ \refcite{IK}.

For $J/\psi$ production all $r^{04}_{ij}$ are consistent with zero, but the errors are quite large. Within the heavy quark approximation all $r^{04}_{ij}$ identically vanish, since only the ${\cal M}_{++}$ amplitude is non-zero. However, the full BFKL calculations predict non-vanishing $r^{04}_{ij}$, and these results are also compatible with the $J/\psi$ data.

\section{Conclusions and outlook}

In summary, the BFKL calculations reproduce the differential cross sections for all vector mesons very well, and also the dominance of transversely polarized mesons and relative importance of the different helicity amplitudes. But they do not correctly account for the sign of the amplitude for longitudinal mesons. The two-gluon calculations, on the other hand, predict a much too large fraction of longitudinally polarized mesons but can despite this describe the differential cross section. However, the problem of endpoint divergences remains.

More theoretical work is clearly needed to understand the process, but in any case it has been demonstrated that BFKL is an important ingredient in the calculations and, importantly, makes the amplitudes finite. The results have also confirmed that the chiral odd components of the photon wave function are important to describe the data. It seems, however, that we may require a larger suppression of large dipoles, and the role of the factorization into a hard process and soft vector meson formation needs to be investigated. Computing the corresponding processes in electroproduction would also give more constraints on the helicity structure.

On the experimental side, it would be helpful to have data for larger momentum transfer and with better statistics. If, with more precise data on $J/\psi$ production, one would see deviations from zero in the spin density matrix elements, it would be a clear sign that the heavy quark approximation is not appropriate and that one should instead use the full framework discussed here.

\section*{Acknowledgments}      
I wish to thank Jeff Forshaw, Leszek Motyka and Gavin Poludniowski, with whom I had a very fruitful collaboration on these topics, and many discussions. I also thank Gunnar Ingelman, Leszek Motyka, Bernard Pire and Lech Szymanowski for discussions and comments which helped improve this paper.


\end{document}